# Retardation of Bulk Water Dynamics by Disaccharide Osmolytes

*Nimesh Shukla[†], E. Pomarico[‡], Lee Chen[†], M. Chergui[‡] and C. M. Othon[†§]\**

[†] Department of Physics, Wesleyan University, Middletown CT 06457 USA

[§] Molecular Biophysics Program, Wesleyan University, Middletown CT 06457 USA

[‡] Laboratoire de Spectroscopie Ultrarapide (LSU) and Lausanne Centre for Ultrafast Science (LACUS), École Polytechnique Fédérale de Lausanne, ISIC, FSB, CH-1015 Lausanne, Switzerland

\*cothon@wesleyan.edu

**ABSTRACT** : The bioprotective nature of disaccharides is hypothesized to derive from the modification of the hydrogen bonding network of water which protects biomolecules through lowered water activity at the protein interface. Using ultrafast fluorescence spectroscopy we measured the relaxation of bulk water dynamics around the induced dipole moment of two fluorescent probes (Lucifer Yellow Ethylenediamine and Tryptophan). Our results indicate a reduction in bulk water reorganization rate of approximately of 30%. We observe this retardation in the low concentration regime measured at 0.1M and 0.25 M, far below the onset of glassy dynamics. This reduction in water activity could be significant in crowded biological systems, contributing to global change in protein energy landscape, resulting in a significant enhancement of protein stability under environmental stress. We observed similar dynamic reduction for two disaccharide osmolytes, sucrose and trehalose, with trehalose being the more effective dynamic reducer.

**TOC GRAPHICS**

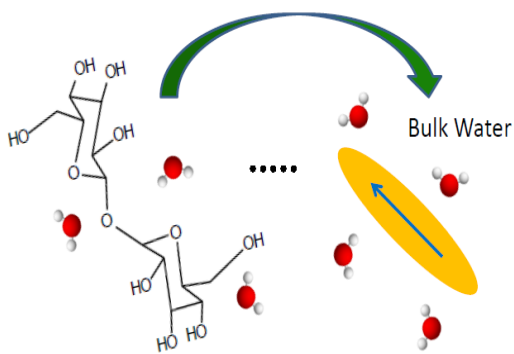

The importance of the physiochemical interactions of solvent molecules in regulating protein and biomolecular structural dynamics cannot be understated. Solvent molecules regulate protein folding and conformational stability by acting as a plasticizer, and it has been demonstrated that the dynamic behavior of a protein follows that of the hydration dynamics of the solvent [1, 2]. The maintenance of protein structure and activity in the cellular environment is carefully controlled by the relative concentration of small molecular osmolytes. These osmoprotectant molecules influence protein structure indirectly by regulating water activity and hydration dynamics. Small molecular solutes that influence water structure and dynamics can be roughly categorized into two groups labelled: kosmotrope and chaotrope. Kosmotrope is a term used to describe molecules that stabilize protein structures through changes in water structure ("structure makers"), whereas chaotropes destabilize protein structures ("structure breakers") [3]. A few examples of kosmotropic molecules are disaccharides, polyols, amino acids, and methylamines.

Disaccharide molecules, which are the focus of the current study, protect protein structures under a range of physical stresses including cryogenic temperature [4, 5], elevated temperature [6-10], dehydration [11-15], and excessive salinity [15-17]. Disaccharides are compatible osmolytes, meaning they do not perturb the native structure of proteins, nor are they limited to a particular structural folding motif or sequence; this makes them particularly useful for pharmaceutical and industrial purposes. Trehalose is the most effective disaccharide cryoprotectant and kosmotrope known. Trehalose is a non-reducing disaccharide of glucose linked through an $\alpha,\alpha\text{-}(1\rightarrow1)$-glycosidic bond, and is more effective at protecting biomolecular structure than other chemically similar disaccharides such as sucrose and maltose. A number of models have been proposed to explain the bioprotective properties of disaccharides. However, a complete picture of the physical mechanism of protection of protein structures remains elusive.

Several hypotheses have been formulated to explain the bioprotective effectiveness of disaccharides including the vitrification, preferential interaction, and water replacement models [18]. The water replacement model applies only to systems under conditions of extreme dehydration. In this model trehalose preferentially solvates biomolecules, thereby replacing water in the hydration layer [18]. There is no evidence for a direct interaction between proteins and disaccharides in the hydrated state, and in fact it has been demonstrated that disaccharides are preferentially excluded from the protein hydration layer [19]. Therefore we do not consider this model relevant to protein structural maintenance living systems under physiological conditions as presently studied. Both the vitrification and preferential solvation models imply that the disaccharides promote protein stability by slowing hydration dynamics and reducing water activity. There is ample experimental data to indicate retarded water dynamics within the hydration layer of disaccharides, including Terahertz spectroscopy [20], Quasi-Elastic Neutron Scattering (QENS) experiments [21], Dynamic light scattering experiments [22, 23], and NMR spectroscopy [24]. The vitrification model however, cannot explain the poor biopreservation properties of highly viscous (high glass transition temperatures) tetrasaccharides which share many of the physiochemical properties of the bioprotective disaccharides [25, 26].

The preferential solvation model attributes the protective properties of disaccharides to strong cosolute-water interactions, which dramatically reduces water-water interaction in the bulk as well as water activity at the surface of the protein. Experimental evidence for preferential solvation exists in the Raman scattering signal for water-disaccharide mixtures [27, 28]. The results demonstrate an exceptional disruption of the hydrogen bonding network for trehalose concentrations above 30 wt %. Accordingly, it is hypothesized that trehalose reorders the bulk hydrogen bonding network more effectively than other solutes, resulting in dramatically slower dynamics and reduced water activity. This work also implies the existence of a critical disaccharide concentration beyond which the tetrahedral hydrogen bond network of water is significantly disrupted. We note however, that the structural enhancement afforded by trehalose for proteins under thermal stress appears linearly dependent upon the sugar concentration, (see Figure 1) to concentrations well below those observed in the Raman scattering work.

The difference between the hydration of various disaccharides has also been probed by molecular dynamics simulation [27-32]. Two studies in particular have compared disaccharides of glucose that differ only in their glycosidic bonds. Choi *et al*. compared 13 different disaccharides and found that trehalose was the most effective dynamic reducer of solvation layer dynamics [29]. They attributed this result to the rigidity of the molecule and the highly anisotropic distribution of water in the hydration layer. Vila Verde *et al.* also observed a superior reduction of water mobility in the solvation layer of trehalose, which they attribute to the topology of the molecule hindering water rotation in the solvation layer [32]. Both groups however, conclude that these effects should only impact dynamics in the first two hydration shells of the disaccharide and thus would not account for enhanced bio-protection by trehalose. Thus, there appears to be a disconnect between the biological enhancement of protein stability and the current measurements of water structure and bulk activity.

Importantly, most previous experiments have probed water dynamics near the disaccharide, but do not attempt to measure hydration dynamics far from the disaccharide surface. The current study compares the hydration of a small probe solutes in various low concentrations of trehalose and sucrose. We use fluorescence frequency up-conversion to measure the relaxation of water around the excited state of two fluorophores: a dye molecule Lucifer yellow ethylenediamine (LYen) and the naturally fluorescent amino acid tryptophan.

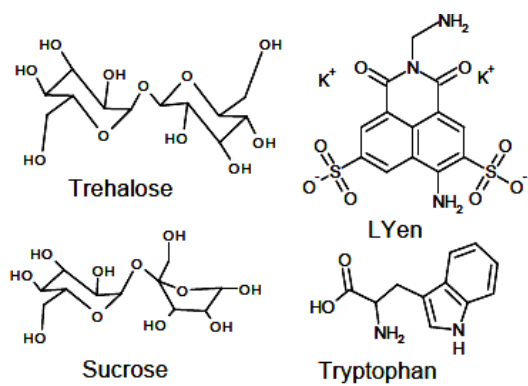

**Schematic 1.** Chemical structure of molecules used in this work.

*Disaccharide Protection Against Thermal Stress.* The linear dependence of the melting temperature as a function of disaccharide concentration has been previously reported for multiple protein models [6-10, 33, 34]. We verified this trend for both sucrose and trehalose using circular dichroism spectroscopy on a model protein system Staph Nuclease (SN). While both trehalose and sucrose stabilize the native structure of protein against thermal stress, trehalose appears to have a slightly larger slope as a function of increasing concentration, see figure 1. Trehalose increases the melting temperature approximately 9 ºC/M, while sucrose gives an increase of approximately 7 ºC/M. If we take the onset of glassy dynamics to be the concentration at which bulk viscosity increases non-linearly, then for both trehalose and sucrose this onset is found at concentrations close to 1.0 M. However, we can see from figure 1 that even though error bars are large there is a significant enhancement in protein stability even at concentration as low as 0.05 M, which is over an order of magnitude lower than that predicted by a vitrification model. We focused our experiments around the low concentration regime to see if there was a notable difference in water mobility, which would be an indicative of dynamic inhibition of the hydrogen bonding network in the bulk caused by these disaccharides.

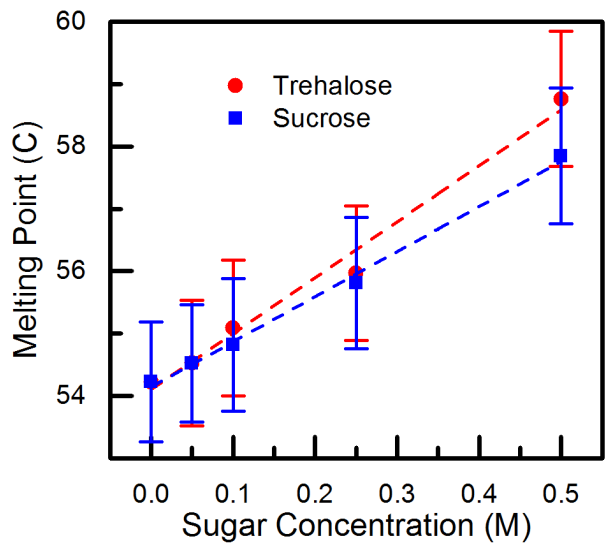

**Figure 1.** Melting temperatures for the enzyme Staph Nuclease as a function of increasing concentration of trehalose and sucrose.

*Measurement of Hydration Dynamics.* Changes in the dynamic activity of water are probed using the dynamic Stokes shift of fluorescent probes. In order to accurately reflect the changes in solvation, the probe spectral properties cannot be impacted by the presence of the disaccharide co-solute. We measured the steady state emission spectra, absorption spectra, and the lifetime of LYen and Tryptophan for disaccharide concentrations of 0M, 0.1M and 0.25M, see figure S1 in supporting information. No difference in emission/absorption spectrum or lifetime was found, this suggests that the disaccharide is not directly interacting with the probe, and any change in the probe dynamics is not due to an alteration of the excited state electronic configuration or solvent exposed area of the probe.

Fluorescence frequency up-conversion was used to measure dynamic changes due to bulk hydration dynamics. Details on the two experimental detection schemes are given in the supplementary information. We measured the fluorescence transients of LYen at ten wavelengths across the spectrum for 0M, 0.1M and 0.25M concentrations of trehalose and sucrose. We observe a fast decay on the blue side of the fluorescence maximum with concomitant rise on the red side of the spectrum indicative of solvation of the excited state [35],figure 2. In order to improve the signal quality, check for sample variation, and eliminate background noise we measured multiple samples and used signal averaging to arrive at the reported hydration values. The use of a lock-in amplifier creates a time average of the signal level at each delay stage position. By collecting the signal intensity as a function of time we collect a fluorescence transient at a particular signal frequency; 20-80 such transients were averaged to eliminate the influence of laser fluctuation and other systematic error. A minimum of two sets of data were taken independently and analyzed in order to insure reproducibility among independent measurements. Finally, the obtained time constants from these data sets were averaged for each concentration. The values obtained for LYen were highly repeatable with fairly low uncertainty in the hydration time constants.

We analyzed our data using global fit (GF) of transients by one Gaussian and three exponentials (two solvation components and one lifetime component) convoluted with the instrument response function, figure 2-a. This form of spectral response has been widely investigated [36-39], and the attribution of these relaxation modes are discussed specifically for LYen by Vauthey *et al*. The results from the global fit for all LYen samples are given in table 1. We are most interested in the solvent hydration which is described by the picosecond relaxation time-scale. The lifetime component of $6.0 \pm 0.1$ ns was independently measured on a Time Correlated Single Photon Counting (TCSPC) apparatus and was fixed during the global fitting for all samples of LYen.

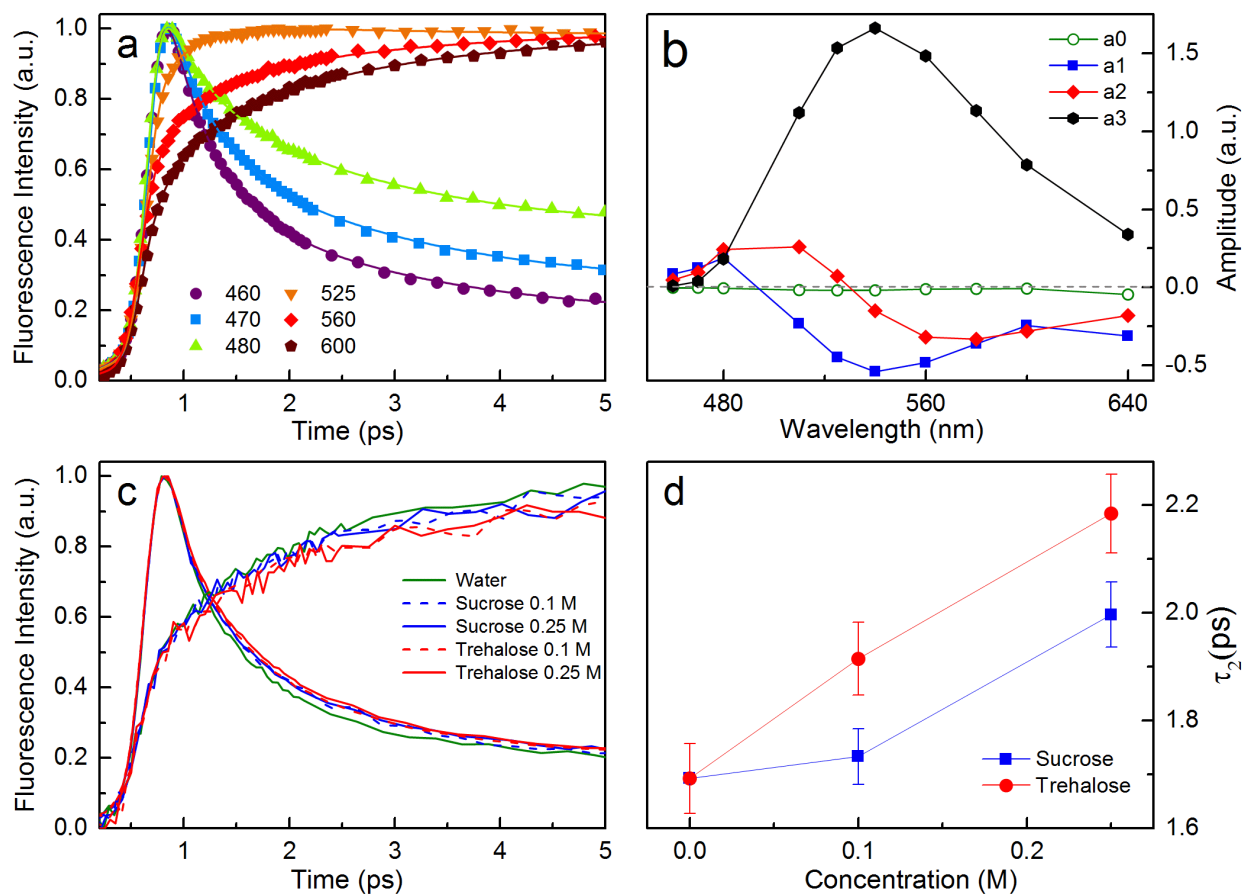

**Figure 2**.a) LYen fluorescence up-conversion transients for wavelengths across the spectrum (points) and their Global fit (solid line) b) Amplitudes for global fit time-constants (a0, a1, a2, a3 are the amplitudes of Gaussian, fs, ps and ns time constants respectively) c) Example up-conversion transients for various concentrations of disaccharides at 460 nm (fast decaying curves) and at 640 nm (slow rising curves). d) Increase in solvation time constant of LYen with respect to disaccharide concentration.

The measured up-conversion transients and their global fits for LYen in pure water are shown in figure 2-a. For LYen in water we observed two solvation components on time scale of ~390fs and ~1.69 ps. The shorter (~fs) component is due to vibration and libration energy loss [35, 40, 41]. The few ps component is due to the solvation of probe by water [35, 40, 41]. From figure 2-b we see the amplitude of the femtosecond and picosecond time constants changes sign from the blue to the red side of spectrum, the signature of solvation [35, 39]. The amplitudes of lifetime component remain positive across the spectrum, as expected. This approach to measuring solvation has been widely applied to analyse changes in water accessibility and hydrogen bond network dynamics [42-45].

Figure 2-c displays representative transients for the probe molecule LYen in the presence of the two disaccharide co-solutes. In presence of the sugars, the solvation of the excited state takes place at a slower rate on both the blue and the red side of the spectral maximum. The difference is small, but detectable across the fluorescence spectrum. We measured a solvation time constant of 1.99±0.04 ps for trehalose and 1.73±0.04 ps for sucrose at a concentration of 0.1 M, and 2.18±0.05 ps for trehalose and 1.91±0.05 ps for sucrose at a concentration of 0.25 M. This is an indicative of distinct retardation of dynamics in the bulk water by ~30% for trehalose and ~15% for sucrose at 0.25M concentration. We observed an increase in solvation time constant with respect to disaccharide concentration as shown in figure 2-d. Trehalose appears to have larger slope than sucrose similar to the melting curve presented for Staph Nuclease, Figure 1; which is consistent with water mobility influencing the stability of protein structures as suggested by preferential solvation model.

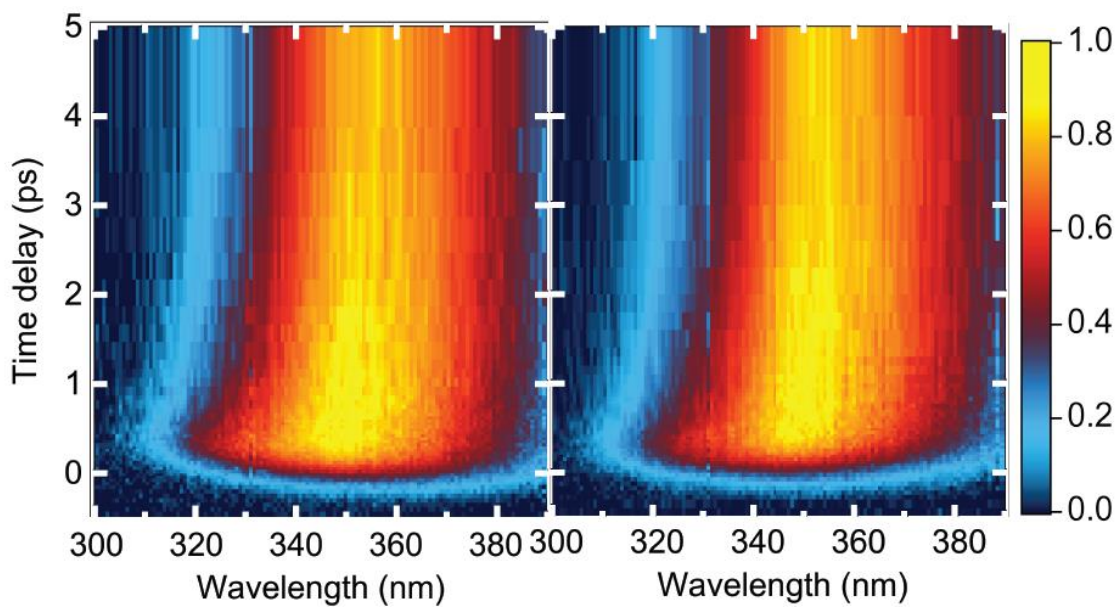

**Figure 3.** Spectral relaxation of tryptophan at 0.1 M concentration of disaccharides a) Trehalose b) Sucrose

As a further exploration of the change in water activity around a solvated biomolecule, we chose to measure the Stokes shift around the naturally fluorescent amino acid tryptophan. This probe has excitation absorption maxima in the ultraviolet range, which is outside the doubled frequency range of the apparatus used for LYen. The full spectral response for both sucrose and trehalose at 0.1 M is shown in figure 3. In order to extract solvation relaxation, the Time Resolved Emission Spectrum (TRES) at each delay time was fit with an asymmetric lognormal function. We extract the first moment of this fit, and use it to monitor the spectral relaxation as a function of time as shown in figure 4.

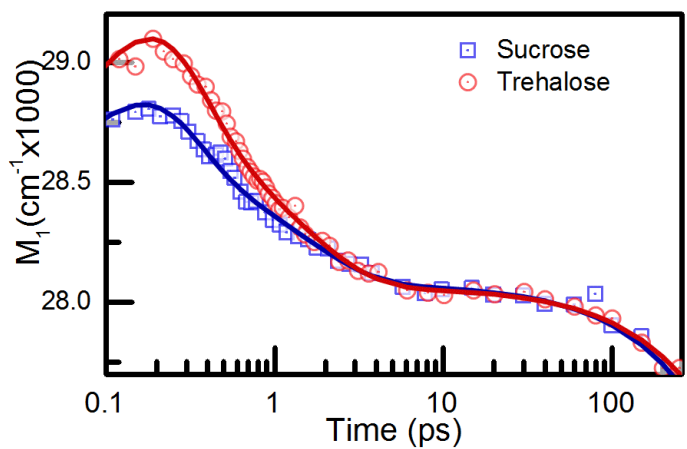

**Figure 4**. Spectral relaxation of tryptophan plotted as function of time for 0.1 M trehalose and 0.1 M sucrose solutions; the error in the spectral fit is comparable to the size of the data point. A three-exponential fit convoluted with Gaussian IRF accurately represents the relaxation as shown (solid lines).

The shift in the spectral maximum of TRES is directly correlated to solvation dynamics of tryptophan. We fit the relaxation curve with a three-exponential function convoluted with a Gaussian IRF of 250 fs in order to extract the solvation relaxation time constants. We use a fixed time constant of 3 ns representing vibration and solvation of water around the probe. Using this method we observed a solvation time constant of 1.4±0.2 ps for trehalose 1.46±0.4 ps for sucrose. Previously, Bräm *et al.* have reported a solvation time constant of 1.02±0.12 ps for tryptophan in pure water [46]. We found the vibration component remains unchanged in the presence of sugars, however the retardation in the hydration relaxation time constant around tryptophan corresponds to a substantial decrease in dynamic activity of water in the bulk. This supports our observation in LYen that both disaccharides were successful in altering the dynamic activity of water in the bulk, far away from disaccharide hydration layer.

Our results significantly extend the range at which retardation of hydration dynamics have to be detected. They complement with work done at much higher concentrations where the hydration layer of cosolute are significantly overlapped. For example, Raman measurements of the water hydrogen bond angle are distorted for sugar concentrations above 1.0 M [27, 28]. More recently the observations made by Sajadi *et al.* [20] suggest that for 0.4 M concentration of trehalose the solvation dynamics in the bulk are identical to those of a terahertz probe which is covalently bound to trehalose. From these results they predict that the retardation of hydration dynamics should extend well beyond the solvation layer of the sugar. Our results support this prediction by showing retardation in bulk hydration dynamics in the low sugar concentration regime for solvated fluorescent probes. At this concentration (0.1M), the average distance between molecules is approximately 21 Å which corresponds to ~7 layers of water [47].

This result is consistent with the categorization of these disaccharides as compatible co-solutes which are preferentially excluded from the protein surface [19]. This study indicates that trehalose is an effective dynamic reducer of bulk water dynamics. Although the large error in solvation time constant in the case of tryptophan restricts our ability to make a distinction between trehalose and sucrose, we observed a slightly stronger concentration dependent retardation of water activity for trehalose than sucrose in case of LYen. Both disaccharides are effective dynamic reducers as well as bio-protective compatible osmolytes. This suggests that the alteration of water activity may be responsible for the linear enhancement in thermal stability for proteins observed for these disaccharides. Large differences in the thermal denaturation become apparent as the concentration is dramatically increased, and therefore the role of viscoelastic and glassy dynamics may contribute significantly to the stabilization of large macromolecules in this concentration range. In conclusion, our results offer an indirect mechanism by which osmolyte molecules such as trehalose stabilise the protein structures by regulating the water activity at the protein surface, while being preferentially excluded from this interface.

|  |  | Lucifer Yellow | | | Tryptophan | |
|---|---|---|---|---|---|---|
|  | Conc. | $\tau_0$ (fs) | $\tau_1$ (fs) | $\tau_2$ (ps) | $\tau_1$ (ps) | $\tau_2$ (ps) |
| Water | 0 | 63.0±0.8 | 390±20 | 1.69±0.05 | 0.16±0.04 * | 1.02±0.12 * |
| Sucrose | 0.1 | 73.4±0.8 | 340±10 | 1.73±0.04 | 0.18±0.1 | 1.46±0.4 |
| Sucrose | 0.25 | 73.2±0.7 | 390±10 | 1.99±0.04 | ----- | ----- |
| Trehalose | 0.1 | 73.1±0.8 | 380±10 | 1.91±0.05 | 0.2±0.2 | 1.4±0.2 |
| Trehalose | 0.25 | 69.8±0.9 | 440±20 | 2.18±0.05 | ----- | ----- |

**Table 1.** Time constants obtained for LYen and Tryptophan in presence of trehalose and sucrose at various concentrations.

*Bräm *et al*.[46]

ACKNOWLEDGEMENTS: We thank Dr. Bertrand Garcia-Moreno E. (Johns Hopkins Department of Biophysics, Baltimore, MD) for generously providing the wild-type Staphylococcal Nuclease construct transformed into BL21(DE3) cells. We thank E. Taylor and her lab for their discussion and technical support. This work has been supported by the Connecticut Space Grant Consortium, and was partially funded by the NCCR:MUST of the Swiss NSF.

SUPPORTING INFORMATION AVAILABLE: Experimental details for the two fluorescence frequency up-conversion apparatus as well as details for the data fitting procedure can be found in the supporting information. In addition steady-state and lifetime data for the two fluorescent probes is provided.

**REFERENCES**

1. Reat, V., et al., *Solvent dependence of dynamic transitions in protein solutions.* Proceedings of the National Academy of Sciences, 2000. **97**(18): p. 9961-9966.
2. Halle, B., *Protein hydration dynamics in solution: a critical survey.* Philos Trans R Soc Lond B Biol Sci, 2004. **359**(1448): p. 1207-23; discussion 1223-4, 1323-8.
3. Yancey, P.H., *Water Stress, Osmolytes and Proteins.* American Zoologist, 2001. **41**(4): p. 699-709.
4. SOMME, L., *Anhydrobiosis and cold tolerance in tardigrades.* EJE, 1996. **93**(3): p. 349-357.
5. Storey, K.B. and J.M. Storey, *Frozen and Alive.* Scientific American, 1990. **263**(6): p. 92-97.
6. Lin, T.Y. and S.N. Timasheff, *On the role of surface tension in the stabilization of globular proteins.* Protein Sci, 1996. **5**(2): p. 372-81.
7. Xie, G. and S.N. Timasheff, *The thermodynamic mechanism of protein stabilization by trehalose.* Biophysical Chemistry, 1997. **64**(1-3): p. 25-43.
8. Carninci, P., et al., *Thermostabilization and thermoactivation of thermolabile enzymes by trehalose and its application for the synthesis of full length cDNA.* Proceedings of the National Academy of Sciences, 1998. **95**(2): p. 520-524.
9. Sola-Penna, M. and J.R. Meyer-Fernandes, *Stabilization against thermal inactivation promoted by sugars on enzyme structure and function: why is trehalose more effective than other sugars?* Arch Biochem Biophys, 1998. **360**(1): p. 10-4.
10. Kaushik, J.K. and R. Bhat, *Why is trehalose an exceptional protein stabilizer? An analysis of the thermal stability of proteins in the presence of the compatible osmolyte trehalose.* J Biol Chem, 2003. **278**(29): p. 26458-65.
11. Wright, J.C., *Cryptobiosis 300 Years on from van Leuwenhoek: What Have We Learned about Tardigrades?* Zoologischer Anzeiger - A Journal of Comparative Zoology, 2001. **240**(3-4): p. 563-582.
12. Madin, K.A.C. and J.H. Crowe, *Anhydrobiosis in nematodes: Carbohydrate and lipid metabolism during dehydration.* Journal of Experimental Zoology, 1975. **193**(3): p. 335-342.
13. Behm, C.A., *The role of trehalose in the physiology of nematodes.* International Journal for Parasitology, 1997. **27**(2): p. 215-229.
14. Crowe, J., *The physiology of cryptobiosis in tardigrades.* Mem. Ist. Ital. Idrobiol., 1975. **32**: p. 37-59.
15. Newman, Y.M., S.G. Ring, and C. Colaco, *The Role of Trehalose and Other Carbohydrates in Biopreservation.* Biotechnology and Genetic Engineering Reviews, 1993. **11**(1): p. 263-294.

16. Purvis, J.E., L.P. Yomano, and L.O. Ingram, *Enhanced trehalose production improves growth of Escherichia coli under osmotic stress.* Appl Environ Microbiol, 2005. **71**(7): p. 3761-9.
17. Hounsa, C.G., et al., *Role of trehalose in survival of Saccharomyces cerevisiae under osmotic stress.* Microbiology, 1998. **144 ( Pt 3)**: p. 671-80.
18. Jain, N.K. and I. Roy, *Effect of trehalose on protein structure.* Protein Sci, 2009. **18**(1): p. 24-36.
19. Timasheff, *The Stabilization of Proteins by Sucrose.* 1981.
20. Sajadi, M., et al., *Observing the Hydration Layer of Trehalose with a Linked Molecular Terahertz Probe.* J Phys Chem Lett, 2014. **5**(11): p. 1845-9.
21. Salvatore Magazu, F.M., † and Mark T. F. Telling‡, *α,α-Trehalose-Water Solutions. VIII. Study of the Diffusive Dynamics of Water by High-Resolution Quasi Elastic Neutron Scattering.* 2005.
22. Gallina, M.E., et al., *Rotational dynamics of trehalose in aqueous solutions studied by depolarized light scattering.* The Journal of Chemical Physics, 2010. **132**(21): p. 214508.
23. Uchida, T., M. Nagayama, and K. Gohara, *Trehalose solution viscosity at low temperatures measured by dynamic light scattering method: Trehalose depresses molecular transportation for ice crystal growth.* Journal of Crystal Growth, 2009. **311**(23-24): p. 4747-4752.
24. Hackel, C., et al., *The trehalose coating effect on the internal protein dynamics.* Phys Chem Chem Phys, 2012. **14**(8): p. 2727-34.
25. Crowe, J.H., J.F. Carpenter, and L.M. Crowe, *The role of vitrification in anhydrobiosis.* Annu Rev Physiol, 1998. **60**: p. 73-103.
26. Fabiana Sussich, et al., *Reversible dehydration of trehalose and anhydrobiosis:from solution state to an exotic crystal?* 2001.
27. Lerbret, A., et al., *Low-frequency vibrational properties of lysozyme in sugar aqueous solutions: a Raman scattering and molecular dynamics simulation study.* J Chem Phys, 2009. **131**(24): p. 245103.
28. Lerbret, A., et al., *Influence of homologous disaccharides on the hydrogen-bond network of water: complementary Raman scattering experiments and molecular dynamics simulations.* Carbohydr Res, 2005. **340**(5): p. 881-7.
29. Choi, Y., et al., *Molecular dynamics simulations of trehalose as a 'dynamic reducer' for solvent water molecules in the hydration shell.* Carbohydr Res, 2006. **341**(8): p. 1020-8.
30. Corradini, D., et al., *Microscopic mechanism of protein cryopreservation in an aqueous solution with trehalose.* Sci Rep, 2013. **3**: p. 1218.
31. Qiang Liu, R.K.S., †,‡ B. Teo,† P. A. Karplus,§ and J. W. Brady*,†, *Molecular Dynamics Studies of the Hydration of α,α-Trehalose.* 1997.
32. Verde, A.V. and R.K. Campen, *Disaccharide topology induces slowdown in local water dynamics.* J Phys Chem B, 2011. **115**(21): p. 7069-84.
33. Chen, L., et al., *Sucralose Destabilization of Protein Structure.* The Journal of Physical Chemistry Letters, 2015. **6**(8): p. 1441-1446.
34. Kumar, A., P. Attri, and P. Venkatesu, *Trehalose protects urea-induced unfolding of alpha-chymotrypsin.* Int J Biol Macromol, 2010. **47**(4): p. 540-5.
35. Maroncelli, J., Fleming,Kumar P., *Femtosecond solvation dynamics in water* nature, 1994. **369**.

36. Eric Vauthey, A.F., *Ultrafast Excited-State Dynamics of Oxazole Yellow DNA Intercalators.* 2007.
37. M. L. Horng, J.A.G., A. Papazyan, and M. Maroncelli, *Subpicosecond Measurements of Polar Solvation Dynamics: Coumarin 153 Revisited.* 1995.
38. Furstenberg, A., et al., *Site-dependent excited-state dynamics of a fluorescent probe bound to avidin and streptavidin.* Chemphyschem, 2009. **10**(9-10): p. 1517-32.
39. Furstenberg, A. and E. Vauthey, *Excited-state dynamics of the fluorescent probe Lucifer Yellow in liquid solutions and in heterogeneous media.* Photochem Photobiol Sci, 2005. **4**(3): p. 260-7.
40. Maroncelli, M. and G.R. Fleming, *Computer simulation of the dynamics of aqueous solvation.* The Journal of Chemical Physics, 1988. **89**(8): p. 5044.
41. Barnett, R.B., U. Landman, and A. Nitzan, *Relaxation dynamics following transition of solvated electrons.* The Journal of Chemical Physics, 1989. **90**(8): p. 4413.
42. Pal, S.K. and A.H. Zewail, *Dynamics of water in biological recognition.* Chem Rev, 2004. **104**(4): p. 2099-123.
43. Kwon, O.H., et al., *Hydration dynamics at fluorinated protein surfaces.* Proc Natl Acad Sci U S A, 2010. **107**(40): p. 17101-6.
44. Othon, C.M., et al., *Solvation in protein (un)folding of melittin tetramer-monomer transition.* Proc Natl Acad Sci U S A, 2009. **106**(31): p. 12593-8.
45. Xu, J. and J.R. Knutson, *Ultrafast fluorescence spectroscopy via upconversion: applications to biophysics*, in *Methods Enzymol*. 2008. p. 159-83.
46. Bram O., O.A., Tortschanoff A.,van Mourik F.,Madrid M.,Echave J.,Cannizzo A.,Chergui M., *Relaxation dynamics of tryptophan in water: A UV fluorescence up-conversion and molecular dynamics study.* J Phys Chem A, 2010. **114**(34): p. 9034-42.
47. Winther, L.R., J. Qvist, and B. Halle, *Hydration and Mobility of Trehalose in Aqueous Solution.* The Journal of Physical Chemistry B, 2012. **116**(30): p. 9196-9207.